\documentclass{webofc}
\usepackage[varg]{txfonts}   % Web of Conferences font
\begin{document}
\title{Bayesian analysis of nontrivial features in the speed of sound inside neutron stars in light of astrophysical and pQCD constraints}

\author{\firstname{Debora} \lastname{Mroczek}\inst{1}\fnsep\thanks{\email{deboram2@illinois.edu}}}

\institute{Illinois Center for Advanced Studies of the Universe, Department of Physics, 
University of Illinois Urbana-Champaign, 1110 W. Green St., Urbana IL 61801-3080, USA.}

\abstract{%
  Functional forms of the neutron star Equation of State (EoS) are required to extract the viable EoS band from neutron star observations. Realistic nuclear EoS, containing deconfined quarks or hyperons, present nontrivial features in the speed of sound such as bumps, kinks, and plateaus. Using modified Gaussian processes to model EoS with nontrivial features, we show in a fully Bayesian analysis incorporating measurements from X-ray sources, gravitational wave observations, and perturbative QCD results that these features are compatible with current constraints. We find nontrivial behavior in the EoS plays a role in understanding the possible phase structure of neutron stars at densities around 2 $n_{\rm sat}$. 
}
\maketitle
\section{Introduction}
\label{intro}
For stable, slowly-rotating neutron stars, structural properties such as the mass, radius, and tidal deformability, are uniquely dependent of the Equation of State (EoS) of cold, beta-equilibrated nuclear matter. EoS calculated from nuclear physics models containing quark and strange degrees of freedom predict multi-scale correlations across baryon density, $n_B$, in the speed of sound functional, $c_s^2(n_B)$, due to the behavior of thermodynamic quantities near phase transitions. In a recent study \cite{Mroczek:2023zxo}, we proposed a new method for modeling multi-scale correlations in the EoS using modified Gaussian processes (mGPs). We performed a Bayesian model comparison of mGP EoS against a benchmark model of EoS containing only long-range correlations. Given mass, radius, tidal deformability, and symmetry energy constraints and perturbative QCD (pQCD) input, we found that the presence of multi-scale correlations in the EoS is neither favored nor disfavored.

\section{Statistical Methods}
Our approach is based on Refs.~\cite{Miller:2019nzo,Mroczek:2023zxo}. The posterior probability of a given EoS, $p_k$, is determined by the product of its prior probability and a likelihood function, normalized by the model evidence. That is, for an EoS $k$ belonging to a model $l$, $p_k$ is
\begin{equation}
    p_k = \dfrac{q_k\mathcal{L}_k}{\int q_l\mathcal{L}_l dl},
\end{equation}
where we define the prior probability $q_k$ to be uniform across all EoS considered. The likelihood of EoS $k$, $\mathcal{L}_k$, given the data, is
\begin{equation}
    \mathcal{L}(\vec{\phi}_k)= \mathcal{L}_{S}(\vec{\phi}_k) \mathcal{L}_{\rm Mmax}(\vec{\phi}_k) \mathcal{L}_{\rm M-R}(\vec{\phi}_k)  \mathcal{L}_{\Lambda}(\vec{\phi}_k)w_{\rm pQCD}(\vec{\phi}_k),
\end{equation}
where $S$ denotes the likelihood factor associated with symmetry energy estimates, $M_{\rm max}$ is associated with high-mass pulsar measurements, $M-R$ corresponds to simultaneous mass-radius measurements, and $\Lambda$ represents tidal deformability data. The input from pQCD is incorporated using a weight, $w_{\rm pQCD}$, as defined in \cite{Gorda:2022jvk}. Lastly, the model evidence for model $l$, $\mathcal{E}_l \equiv {\int q_l\mathcal{L}_l dl}$, is the integral over the posterior probabilities for all EoS from model $l$ and it quantifies the level of support of the data for the model. In Bayesian model comparison, we use the Bayes factor, $K = \mathcal{E}_l/\mathcal{E}_m$, to quantify the level of support of the data for model $l$ over a competing model $m$. A significant deviation from unity indicates a preference for one model over the other. 

\subsection{Equation of state priors}

Approaches based on Gaussian processes \cite{Landry:2018prl,Essick:2019ldf} have become common in the literature as a tool for generating EoS priors that does not make assumptions about system composition and interactions. In these frameworks, the EoS is modeled via an auxiliary variable $\phi_k\equiv\log(1/c_{s,k}^2 - 1)$, where $\phi_k$ is sampled from $\Phi(x_i) \sim \mathcal{N}(\mu_i,\Sigma_{ij})$. Here, $\Phi$ is a collection of functions, normally distributed at each domain point $x_i$ around a mean $\mu_i$. A covariance kernel specifies the covariance matrix $\Sigma_{ij}$. A common choice is the squared-exponential kernel, 
\begin{equation}
    K_{\rm se}\left(x_i,x_j\right) = \sigma^2\exp{-\left(x_i-x_j\right)^2/2\ell^2}.
\end{equation}
In this case, $\sigma$ determines the strength of the correlation between $\Phi(x_i)$ and $\Phi(x_j)$, and $\ell$ the correlation length, such that, if $x_i - x_j\gg \ell$, $\Phi(x_i)$ and $\Phi(x_j)$ are uncorrelated. 

\begin{figure}
   \centering
\begin{tabular}{cc}
\includegraphics[width=0.47\linewidth]{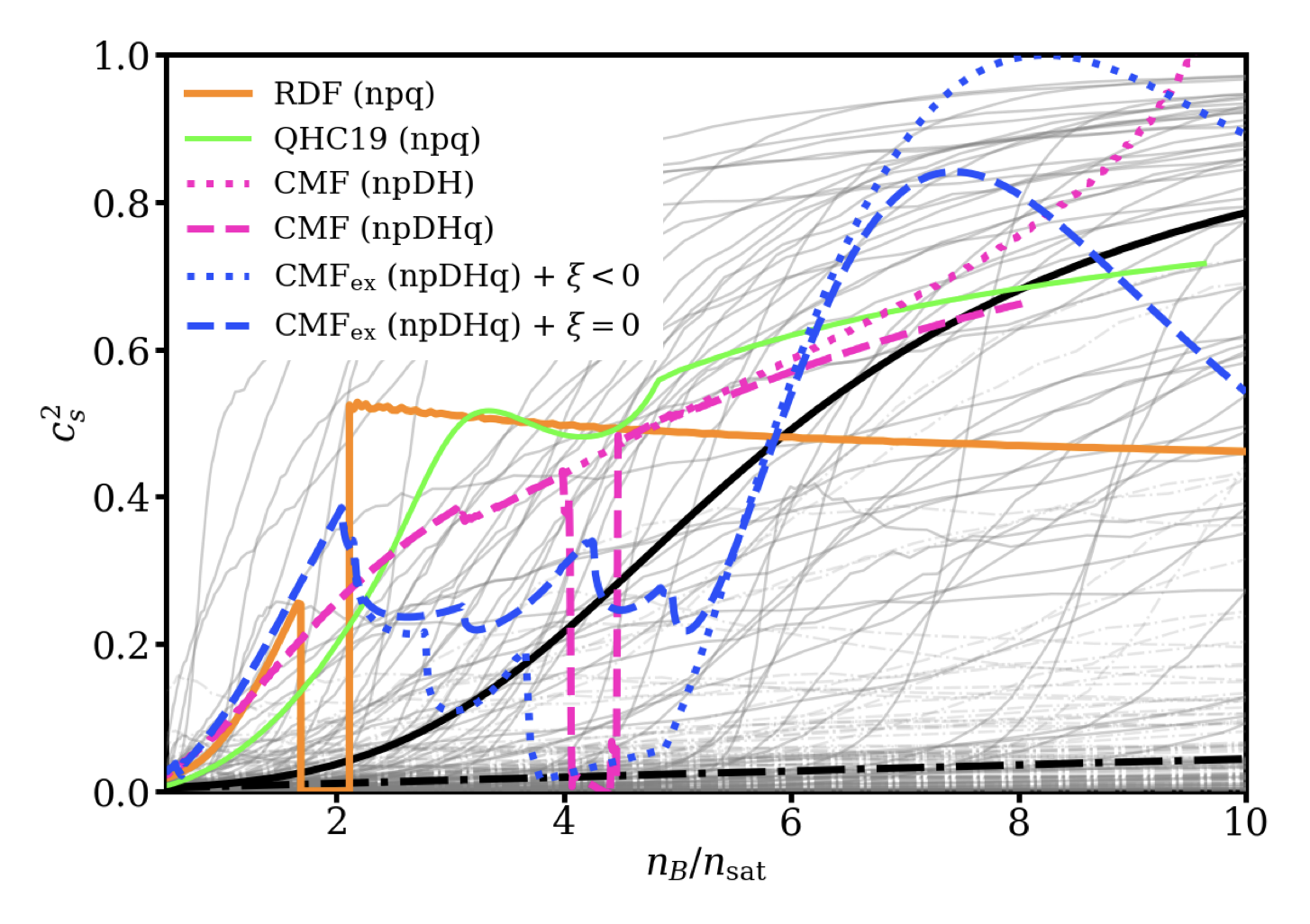} & 
\includegraphics[width=0.47\linewidth]{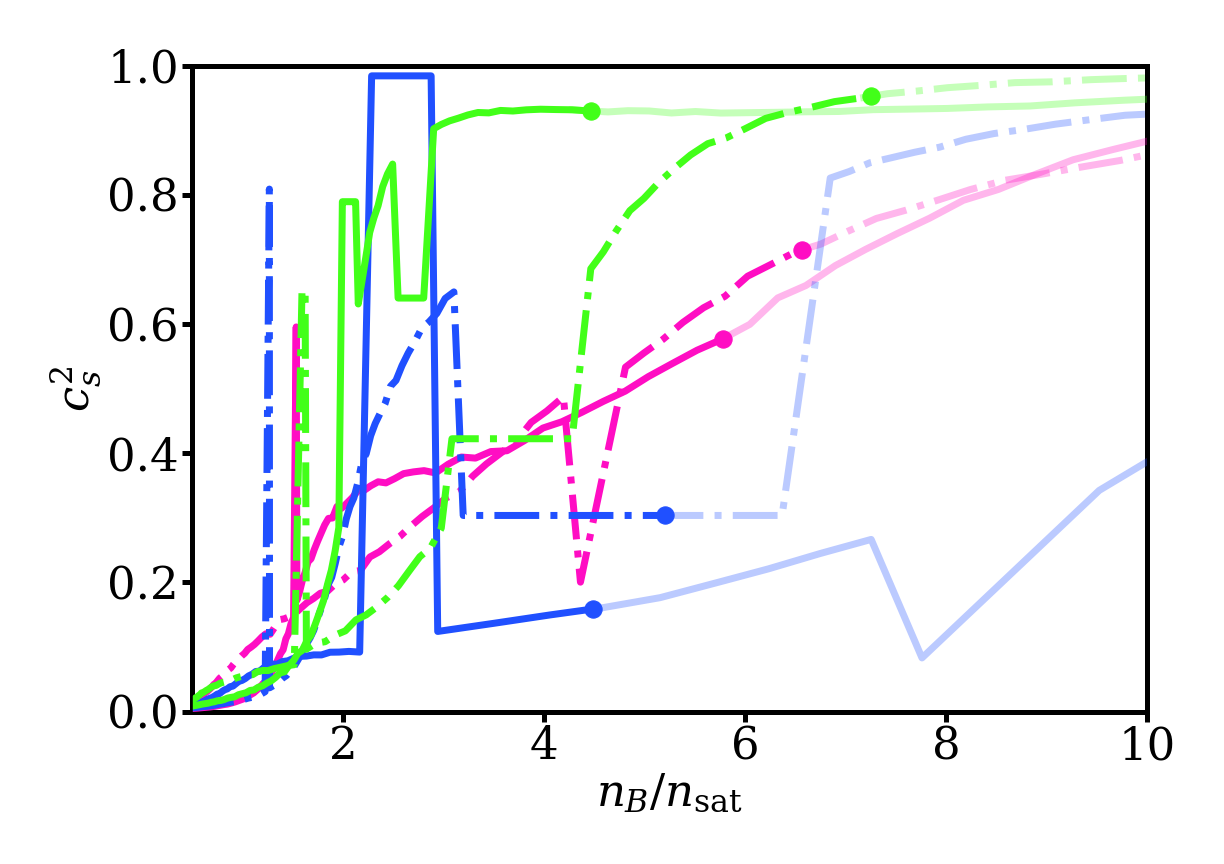}
\\

\end{tabular}
    \caption{Left: The speed of sound as a function of baryon number density in units of $n_{\rm sat}$ for a collection of microphysical models (see \cite{Mroczek:2023zxo} for references) which include nucleons (np), delta resonances (D), hyperons (H), quarks (q), and leptons. In light gray, a sample of functional forms from the benchmark GPs. The black, dot-dashed and solid lines, represent the two means used in the GPs. The benchmark GPs capture a wide range of behavior, but the sharp and nontrivial features in $c_s^2$ observed in state-of-the-art nuclear physics simulations are exponentially suppressed.
    Right: The speed of sound squared in units of $c^2$ as a function of the baryon number density in units of $n_{\rm sat}$ for a representative set of samples generated using the mGP framework. The circles in the middle panel represent the maximal central density predicted for a stable, slowly-rotating star. The mGP framework produces a diverse set of EoS which contain multi-scale correlations across densities.}
    \label{fig:model}
\end{figure}

We define two models for the EoS: a benchmark model, which contains EoS drawn from a standard GP with long-range correlations across densities only, and the mGP model, for which every EoS contains multi-scale correlations. In the mGP, multi-scale behavior is introduced via piecewise modifications in the forms of spikes and plateaus to functional forms of $c_s^2(n_B)$ sampled from the benchmark model. The left panel in Fig.~\ref{fig:model} shows $c_s^2(n_B/n_{\rm sat})$, where $n_{\rm sat} \equiv 0.16$ fm$^{-3}$ is the nuclear saturation density, for several nuclear physics models in different colors. In gray, samples from the benchmark model illustrate that it fails to capture short- and medium-range correlations seen in the nuclear physics simulations. The right panel in Fig.~\ref{fig:model} shows $c_s^2(n_B/n_{\rm sat})$ for a few representative mGP samples and demonstrates that the mGP reproduces the features seen in nuclear physics models with exotic degrees of freedom. 

%%%%%%%%%%%%%%%%%%%%%%%%%%%%%%%%%%%%%%%%%%%%%%%%%%%%%
%%%%%%%%%%%%%%%%%%%%%%%%%%%%%%%%%%%%%%   

%----------------------------------------

\subsection{Astrophysical and theoretical constraints}
As maximum mass constraints, we consider the highest reliably measured neutron star masses \cite{demorest2010two,arzoumanian2018nanograv,antoniadis2013massive}. We also include tidal deformability constraints from GW170817 \cite{LIGOScientific:2017vwq,De:2018uhw,LIGOScientific:2018cki} and GW190425 \cite{LIGOScientific:2020aai}, and the simultaneous mass and radius measurements for PSR J0740+6620 \cite{Miller:2019cac} and PSR J0030+0451 \cite{Miller:2021qha}. From low-energy nuclear experiments, we impose an estimated symmetry energy of $32\pm2$ MeV \cite{Tsang:2012se}. We incorporate partial N3LO pQCD results propagated down to neutron star densities using causality, stability, and thermodynamic consistency \cite{Gorda:2022jvk}. 

%%%%%%%%%%%%%%%%%%%%%%%%%%%%%%%%%%%%%%%%%%%%%%%%%%%%%
\begin{figure}
\centering
\sidecaption
\includegraphics[width=6cm,clip]{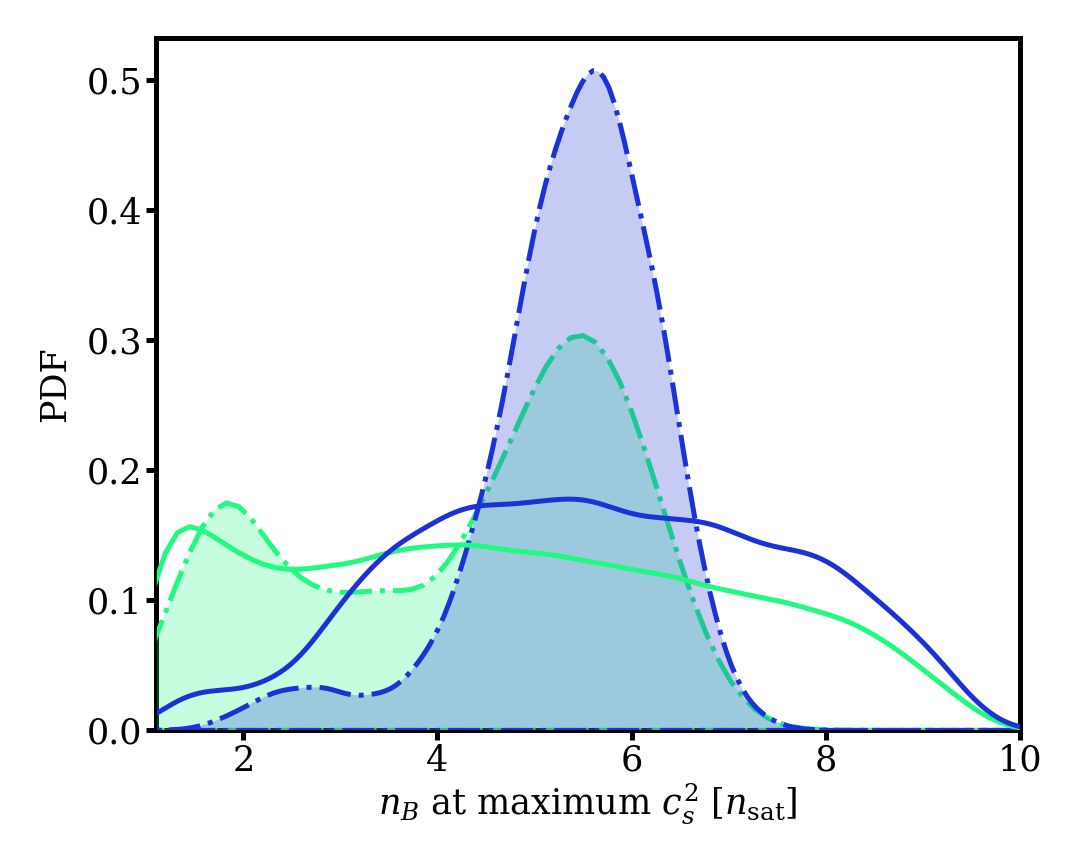}
\caption{Estimated prior probability density distributions (solid curves) and estimated posterior probability density distribution (dot-dashed, filled curves) for the location of a global maximum in $c_s^2(n_B)$ in units of $n_{\rm sat}$ after astrophysical and theoretical constraints are imposed. A global $c_s^2$ maximum at densities below 3 $n_{\rm sat}$ is not ruled out by data, which are also not yet informative enough to favor or disfavor it over a global $c_s^2$ maximum above 3 $n_{\rm sat}$.}
\label{fig-2}       % Give a unique label
\end{figure}
%%%%%%%%%%%%%%%%%%%%%%%%%%%%%%%%%%%%%%   

%----------------------------------------

\section{Results}
The physically motivated long-, medium-, and short-range correlations in the functional form of $c_s^2(n_B)$ in the mGP EoS allow us to model nontrivial structure in $c_s^2(n_B)$. In order to verify if the data more strongly supports such structure over long-range correlations only, we calculated the Bayes factor between the mGP and benchmark EoS. 
We found a Bayes factor $K = \mathcal{E}_{\rm benchmark}/\mathcal{E}_{\rm mGP} = 1.5$,
which suggests that current constraints do not favor either model. 

Furthermore, a bump in $c_s^2(n_B)$ signaling a softening of the EoS has been proposed as a signature of a crossover transition to new, possibly quark, degrees of freedom. Such feature would imply a global maximum in $c_s^2(n_B)$ that occurs within densities realized in stable neutron stars. Figure \ref{fig-2} shows the prior and posterior probability density distributions of value of $n_B$ at the global maximum of $c_s^2$. The prior curves are shown in solid lines for the benchmark EoS (blue) and the mGP EoS (green). In the priors, we see that the multi-scale correlations present in the mGP samples allow for a bump in $c_s^2(n_B)$ before 3 $n_{\rm sat}$. That is not the case in the benchmark model, for which long-range correlations with the regime in which $c_s^2$ must be small to compatible with low-energy nuclear experiments lead to a suppression of large values of $c_s^2$ below 3 $n_{\rm sat}$. 

The posterior curves (dot-dashed, filled) show that in the benchmark model a bump in $c_s^2(n_B)$ is disfavored because the posterior peaks in the range of $n_B$ that characterizes the maximum central density that can achieved in neutron stars for most models, $4-7$ $n_{\rm sat}$. On the other hand, the mGP posterior is bimodal, suggesting that data supports a bump in $c_s^2(n_B)$ before 3 $n_{\rm sat}$. This behavior is compatible with a crossover phase transition in neutron stars. 

\section{Discussion}

Nuclear physics models predict nontrivial features in $c_s^2(n_B)$ and multi-scale correlations across densities when exotic degrees of freedom are present. We introduced modified Gaussian processes as a phenomenological approach to model nontrivial features in $c_s^2(n_B)$. A fully Bayesian analysis including astrophysical, low-energy, and pQCD constraints showed that nontrivial features in the $c_s^2(n_B)$ inside neutron stars are not ruled out, but neither are they required. An analysis of signals of a crossover phase transition demonstrated that multi-scale correlations are important for searching for a crossover transition inside neutron stars, but no preference was observed for a crossover scenario over other possibilities. 

A definitive ruling on the interior composition of neutron stars will require more precise input from astronomical observations, nuclear physics experiments, and effective theories/QCD at high densities. Fortunarely, more data is anticipated from the NICER and LIGO/Virgo/KAGRA collaborations, as well as new nuclear physics measurements at the Facility for Rare Isotope Beams at low densities, and low-energy heavy-ion collisions that will probe the high-density, low-temperature regime of the QCD phase diagram.

\bibliography{inspire}
\end{document}